\documentclass[letterpaper]{article} 
\usepackage{aaai25}  
\usepackage{times}  
\usepackage{helvet}  
\usepackage{courier}  
\usepackage[hyphens]{url}  
\usepackage{graphicx} 
\usepackage{natbib}  
\usepackage{caption} 
\usepackage{algorithm}

\usepackage{newfloat}
\usepackage{listings}
\DeclareCaptionStyle{ruled}{labelfont=normalfont,labelsep=colon,strut=off} 
\floatstyle{ruled}
\newfloat{listing}{tb}{lst}{}
\floatname{listing}{Listing}
\usepackage{booktabs} 
\usepackage{subcaption}
\usepackage[utf8]{inputenc} 
\usepackage{url}            
\usepackage{booktabs}       
\usepackage{amsfonts}       
\usepackage{nicefrac}       
\usepackage{microtype}      
\usepackage{amsthm}
\usepackage{algorithm,algpseudocode}
\RequirePackage{amsmath, subcaption}
\RequirePackage{amsmath,dsfont}
\usepackage{amssymb,accents}
\usepackage{amssymb}
\usepackage[font=small,labelfont=bf]{caption}
\usepackage{multirow}

\usepackage{caption}
\input{Definitions}

\title{Weak Supervision for Improved Precision in Search Systems}

\author{Sriram Vasudevan}
\affiliations{
  LinkedIn Corporation\\
  USA\\
  svasudevan@linkedin.com
}

\begin{document}

\maketitle

\begin{abstract}
Labeled datasets are essential for modern search engines, which increasingly rely on supervised learning methods like Learning to Rank and massive amounts of data to power deep learning models. However, creating these datasets is both time-consuming and costly, leading to the common use of user click and activity logs as proxies for relevance. In this paper, we present a weak supervision approach to infer the quality of query-document pairs and apply it within a Learning to Rank framework to enhance the precision of a large-scale search system.
\end{abstract}

\section{Introduction}
\label{sec:intro}
Industrial search systems that leverage supervised learning and deep learning techniques require large volumes of high-quality labeled data to produce relevant results. One of the key challenges in developing these systems is the significant time and cost involved in manually labeling massive datasets. This process often requires training Subject Matter Experts (SMEs), providing comprehensive guidelines, and waiting several months to curate a meaningful volume of graded relevance labels. Compounding this challenge is the fact that such data can quickly become outdated, necessitating repeated annotation efforts.

To circumvent the costs of creating ``golden'' datasets, search and recommendation systems frequently rely on user activity logs as implicit labels for user-query-document interactions. These logs treat user actions on previously displayed results as feedback on relevance. While this approach helps address data scarcity, it often causes search engines to optimize for engagement rather than true relevance. Although engagement and relevance are correlated, models trained solely on activity logs may exhibit the Matthew Effect \cite{perc2014matthew}, amplify clickbait, and over-rely on activity-based features. This correlation can further break down in cases where user interface signals are ambiguous. For instance, a ``dismiss'' button might indicate disinterest, a temporary lack of relevance, or simply a desire to clear viewed results. As a result, engagement-optimized models can suffer from reduced precision and recall.

To address these issues, the industry has increasingly explored \textit{weak supervision}, a set of methods for generating noisy yet informative training labels efficiently and at scale. Early approaches utilized curated data sources \cite{mintz2009distant} or aggregated crowdsourced labels \cite{dalvi2013aggregating}. More recently, Snorkel \cite{ratner2017snorkel} introduced the idea of SMEs authoring multiple heuristics, or labeling functions (LFs), with varying accuracies and coverage, which are then aggregated into a single label per data point \cite{ratner2016data, bach2017learning}. Snorkel Drybell \cite{bach2019snorkel} extended this concept by incorporating organizational knowledge to refine heuristics and improving scalability through sampling-free aggregation techniques. The rise of Large Language Models (LLMs) further enhances weak supervision, with LLMs now being used as powerful heuristics themselves \cite{hsieh2023distilling, kojima2022large}.

However, a limitation of existing aggregation approaches is their focus on achieving consensus among heuristics without explicitly optimizing for label accuracy, often due to the absence of ground truth data. However, a more common scenario in industrial settings is the availability of a small dataset of ground truth labels obtained through human annotation, albeit insufficient to train Deep Neural Networks (DNNs) at scale. This scenario presents an opportunity to combine organizational knowledge with a limited ``golden labeled dataset'' to simplify heuristic aggregation, thereby scaling up weak supervision while minimizing noise.

In this paper, we describe a distributed, scalable weak supervision solution that we successfully deployed in production to significantly improve the precision of a large-scale job search system. Building upon Snorkel's programmatic approach, we propose a novel technique that leverages SME-authored heuristics, enriched with a seed set of ground truth labels, to generate high-quality training data at scale.

\section{Related Work}\label{sec:related}
Snorkel \cite{bach2017learning, ratner2016data, ratner2017snorkel} is a weakly supervised ML framework that allows SMEs to programmatically label datasets using rules or heuristics, known as Labeling Functions (LFs), with varying accuracy and coverage. LFs can output multi-class labels, abstain, and may also be correlated. Snorkel combines LF outputs using a sampling-based, unsupervised generative model that learns from the agreements and disagreements among LFs, without requiring labeled data. This approach assumes that LFs meet minimum thresholds for accuracy and coverage. The resulting ``consensus model'' generates probabilistic labels for a much larger unlabeled dataset, which is then used to train a discriminative classifier, enabling supervised learning without ground truth labels.

Snorkel Drybell \cite{bach2019snorkel} adapts Snorkel for industrial-scale deployment, addressing challenges like scalability and reliance on handcrafted LFs. It scales to large datasets by adopting a distributed computation backend and replacing the sampling-based generative model with a more efficient, sampling-free approach implemented in TensorFlow \cite{abadi2016tensorflow}. To reduce reliance on manual LFs, Drybell introduces a template-driven interface that integrates existing organizational knowledge, such as internal models and taggers, into the labeling process.

In \cite{nitzan1982optimal}, the authors demonstrate that weighted majority voting is the optimal decision rule for aggregating the decisions of $m$ voters (under certain assumptions). \cite{berend2014consistency} further refines this result, showing that the rule holds only when high-confidence (frequentist) weight estimates are available.

In this work, we build on these foundations by leveraging organizational knowledge bases to streamline LF creation and improve labeling accuracy. We adopt a probabilistic model trained on a small annotated dataset with binary outcomes, a common resource in industrial settings. This approach leads to a simpler, scalable labeling model (equivalent to a weighted majority voter) and improves weak labeling accuracy. Our system operates at scale, labeling hundreds of millions of data points efficiently.
\section{System Architecture}\label{sec:system}

\begin{figure}[htbp]
	\centering
	\includegraphics[width=0.9\linewidth]{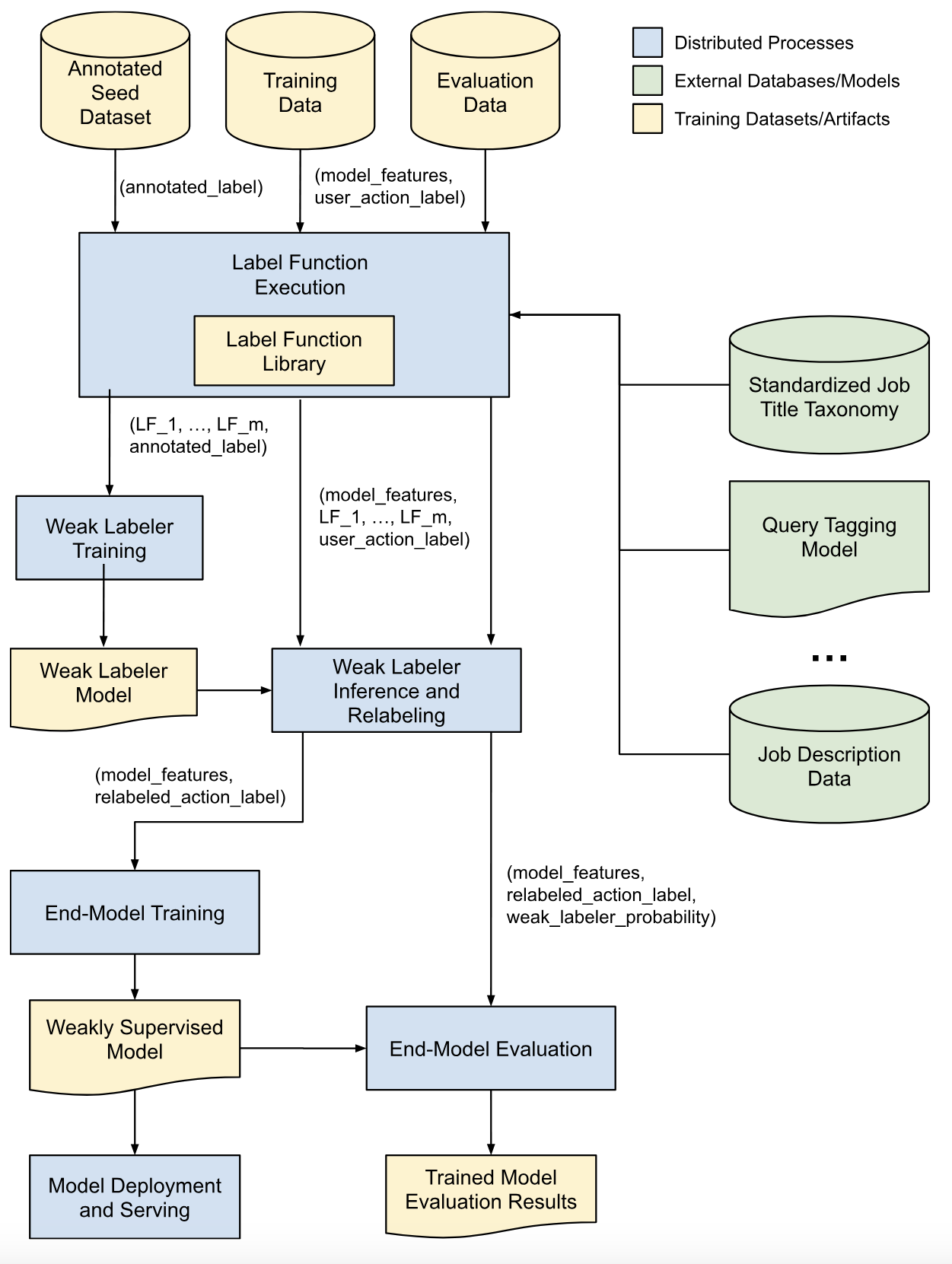}
	\caption{End-to-end design of the weak supervision system and its interaction with external data sources. The distributed processes are built using Apache Spark and TensorFlow.}
	\label{fig:architecture}
\end{figure}

\subsection{Label Function Evaluation}
This stage is implemented using Apache Spark \cite{armbrust2015spark, zaharia2016apache}, which processes the annotated dataset alongside the end-model’s training and evaluation datasets to execute a set of $m$ Label Functions on each record. The LFs are implemented as Spark User Defined Functions (UDFs) for scalability, and they leverage external databases, models and taxonomies to make heuristic decisions. For example, Figure~\ref{fig:architecture} shows the system leveraging a standardized taxonomy reference, a machine learning model and an external database during LF execution.

Unlike Snorkel, which supports multi-class labels, our solution focuses on binary labels for the annotated data (though the end-model's training dataset may be multi-class). Each LF outputs \textit{True}, \textit{False}, or \textit{null}, representing a positive vote, negative vote, or abstention.

If an LF meets latency requirements and avoids using future information (e.g., downstream conversion signals), it can also be served online. In such cases, the LF is added as a feature to the end-model, directly enhancing its performance.

\subsection{Weak Labeler Training, Inference and Relabeling}
The weak labeler aggregates LF outputs into a single probabilistic label, formulated as a supervised learning task. We train a generative model on a small annotated seed dataset, treating LF outputs as features. Due to its simplicity (as described in Section~\ref{sec:weak}), the model is efficiently implemented in Apache Spark, which also handles scoring the weak labeler on the end-model's training and evaluation datasets.

Snorkel’s generative model focuses on minimizing LF disagreements without relying on labeled data, effectively acting as a ``consensus model''. In contrast, our approach leverages the annotated dataset to weigh LFs based on discriminative power, improving labeling accuracy.

We use the weak labeler’s output probabilities to relabel the end-model’s training and evaluation datasets, replacing $y$ with $E_p[y]$, where $p$ is the weak labeler’s output (see Section~\ref{sec:relabel-weak} for details). Unlike Snorkel, which trains the end-model directly on $p$ due to the absence of ground truth labels, our method refines the existing labels, $y$.

\subsection{End-Model Training and Serving}
The end-model is a Deep Neural Network (DNN) with tens of millions of parameters, trained on hundreds of millions of data points. It is trained in a distributed environment using TensorFlow \cite{abadi2016tensorflow} and Horovod \cite{sergeev2018horovod}.

The trained model is deployed online using TensorFlow Serving \cite{tfserving2016}. Any LFs identified as ``serveable'' are incorporated as additional input features during both training and serving.

\subsection{End-Model Evaluation}\label{sec:eval}
The evaluation dataset is scored and relabeled by the weak labeler, following the same process as the training data (Section~\ref{sec:relabel}). We compute NDCG@k on three sets of labels: (1) the original labels, to measure performance on the initial engagement task; (2) the updated labels, to evaluate improvements from weak supervision; and (3) the weak labeler’s predictions, to gauge how well the end-model has learned from the weak labels. Model evaluation is implemented in Spark.

\section{Model Design}\label{sec:model}

\subsection{Weak Labeling Model}\label{sec:weak}
To aggregate the LF ``votes'' into a single probability score, we frame the problem as a supervised learning task where the LF outputs serve as input features, and a small, binary-annotated dataset provides ground truth labels. This annotated dataset is significantly smaller than the end-model’s training data, addressing the challenge of scaling human annotations. Given the limited number of features (LFs) and the small dataset size, we opt for a low-complexity probabilistic generative model to avoid overfitting. To further simplify the model, we assume the LFs are independent.

Let $m$ represent the number of Labeling Functions, $y$ be the true label, and $z_i$ denote the output of the $i\textsuperscript{th}$ LF. The labels are defined as $y \in \{0, 1\}$ for negative and positive classes, respectively, and $z_i \in \{0, 1, \phi\}$, where $\phi$ indicates abstention. The log-odds can be expressed as:
\begin{equation}\begin{split}\label{eq:1}
\log\left(\frac{p}{1-p}\right) &= \log\left(\frac{P(y=1|Z)}{P(y=0|Z)}\right)\\
&=\log\left(\frac{P(Z|y=1)}{P(Z|y=0)}\cdot\frac{P(y=1)}{P(y=0)}\right)
\end{split}\end{equation}

Assuming independence among LFs, Equation~\ref{eq:1} simplifies to:
\begin{equation}
\begin{split}
\label{eq:2}
\log\left(\frac{p}{1-p}\right) &= \log\left(\prod_{i=1}^m \frac{P(z_i|y=1)}{P(z_i|y=0)}\cdot\frac{P(y=1)}{P(y=0)}\right)\\
&= \sum_{i=1}^m \log\left(\frac{P(z_i|y=1)}{P(z_i|y=0)}\right) \\
&\quad + \log\left(\frac{P(y=1)}{P(y=0)}\right)
\end{split}
\end{equation}

For each $z_i$, we define three binary features $x_{ia} = \mathds{1}_a(z_i)$, where $\mathds{1}_a(x)$ is the indicator function for $a \in \{0, 1, \phi\}$. This allows us to rewrite Equation~\ref{eq:2} as a weighted linear model $\text{logit}(p) = w^T x + b$ where the weights and bias are defined as:
$$w_{ia}=\log\left(\frac{P(z_i=a|y=1)}{P(z_i=a|y=0)}\right),\;\; b=\log\left(\frac{P(y=1)}{P(y=0)}\right)$$

The probabilities $P(y)$ and $P(z_i = a | y)$ are estimated from the annotated dataset. This formulation effectively reduces to weighted majority voting, allowing for efficient coefficient estimation and probability computation using standard distributed computing frameworks.

While the independence assumption theoretically complicates LF design (requiring each LF to avoid capturing correlated signals), in practice, minor violations of this assumption do not significantly impact the final ranking model’s performance. This is because the weak labeler’s outputs are treated as inherently noisy.

\subsubsection{Annotated Dataset Size}
To estimate the required dataset size, we assume the LFs have binary outcomes (no abstentions) and model each LF as a Bernoulli process over $n$ records. Assuming the estimation error follows a normal distribution, we have
$p \approx \hat{p} \pm z_\alpha\sqrt{\frac{\hat{p}(1-\hat{p})}{n}}$
where $\hat{p} = k/n$, with $k$ representing the number of times the LF outputs $1$. For a $95\%$ confidence interval, $z_\alpha = 1.96$, leading to a maximum error bound of $\pm 2\sqrt{\frac{0.5\cdot0.5}{n}}$. This implies that to achieve an error less than $E$, we require approximately $1/E^2$ samples. For example, achieving an error of $\leq 5\%$ would need about $400$ samples.

Therefore, the number of samples required to reliably estimate each LF's output falls in the range of hundreds to thousands -- significantly fewer than the hundreds of millions of data points typically needed to train a large DNN model. This makes curating the golden dataset relatively simple, requiring only a few hours of work from an in-house annotation team to produce reliable and accurate labels.

\subsection{Using Weak Labels for Model Training}\label{sec:relabel-weak}
Our job search ranking model uses a \textit{listwise} Learning to Rank approach. This is typically a better choice for ranking tasks because it deals with the relative ordering of items rather than modeling absolute \textit{pointwise} scores. Specifically, our model optimizes ListNet \cite{cao2007learning} or the listwise softmax cross-entropy loss \cite{TensorflowRankingKDD2019}:

\begin{equation}\label{eq:3} \hat{L}(Q, D) = -\frac{1}{q}\sum_{i=1}^q\sum_{j=1}^{n_i} y_{i,j}\cdot \log\left( \frac{\exp{(\hat{y}_{i,j})}}{\sum_{k=1}^{n_i}\exp{(\hat{y}_{i,k})}}\right)\end{equation}
where $q$ is the number of queries, $n_i$ is the number of documents $D_j$ for each query $Q_i$, and $y_{i,j}$ and $\hat{y}_{i,j}$ are the target relevance value and predicted score respectively.

Our weak labeler predicts the probability $p$ of a job being ``extremely irrelevant'' (false positive), to improve the ranking model's precision. This is incorporated into Equation \ref{eq:3}:
\begin{equation}\begin{split}
\hat{L}(Q, D) &= -\frac{1}{q}\sum_{i=1}^q\sum_{j=1}^{n_i} (1-p)\cdot y_{i,j}\cdot \log\left( \frac{\exp{(\hat{y}_{i,j})}}{\sum_{k=1}^{n_i}\exp{(\hat{y}_{i,k})}}\right)\\
&+ p\cdot y_p\cdot \log\left( \frac{\exp{(\hat{y}_{i,j})}}{\sum_{k=1}^{n_i}\exp{(\hat{y}_{i,k})}}\right)
\end{split}\end{equation}
\begin{equation}\label{eq:5}
=-\frac{1}{q}\sum_{i=1}^q\sum_{j=1}^{n_i} \left[(1-p)\cdot y_{i,j} + p\cdot y_p\right]\cdot \log\left( \frac{\exp{(\hat{y}_{i,j})}}{\sum_{k=1}^{n_i}\exp{(\hat{y}_{i,k})}}\right)
\end{equation}

Here, $y_p$ represents the label that would be assigned if $y_i$ were identified as a false positive. Notice that the weakly supervised loss simplifies to merely updating the ground truth labels, eliminating the need for any model or loss modifications. This reduction is broadly applicable whenever the label terms can be factored out of the loss function. Unlike Snorkel, which lacks ground truth labels, our approach leverages weak supervision as a prior or regularizer to fine-tune the target model's performance.

\section{Experiments and Results}\label{sec:experiment}
\subsection{Seed Dataset Preparation}
To develop our weak supervision system, we curated a golden dataset by sampling approximately 1500 representative queries from the search logs and choosing the top 3 documents for each, focusing on improving the precision of the top $k$ results. The resulting 4500 (user, query, document) triplets were annotated by an in-house team, labeling each document as either ``extremely irrelevant'' or not. The annotation task was framed using negation, with the primary goal of reducing egregiously poor results, while still improving overall search precision.

\subsection{Label Function Creation}
We created 10 LFs to determine whether a retrieved document was relevant to the provided implicit and explicit context. Examples include:
\begin{itemize}
\item If the query contains a job title, the search tokens must appear in the title of the retrieved job.
\item The seniority difference between the user and the retrieved job should not exceed one level (seniorities are predicted by another model).
\item If the query includes a job title, its industry must match that of the job (title-industry relationships are defined in a taxonomy file).
\end{itemize}
Some LFs relied on simple string matching, while others leveraged models, databases, and taxonomies to make decisions. As noted earlier, the LF outputs are in $\{0, 1, \phi\}$.

\subsection{Updating Labels of the Training Dataset}\label{sec:relabel}
The search ranking model is trained on user activity logs, with different label values $y_i$ being assigned to different interactions. For example, a user clicking on a result and applying to that job might have the highest value while a user dismissing the job might be given the lowest value. We tried the following relabeling techniques using the weak labeler's output probabilities, in Equation \ref{eq:5}:
\begin{enumerate}
\item \textbf{R1:} $y_p = y_{dismiss}$
\item \textbf{R2:} $y_p = 0$
\item \textbf{R3:} $y_p = y_{dismiss}$ for organic; $p = 0$ for advertised jobs.
\end{enumerate}

\subsection{Offline Results}
\subsubsection{Weak Labeler Validation}
We evaluated the weak labeler by splitting the 4,500-record golden dataset into an 80-20 train-test split. The generative model achieved an AUC of \textbf{0.86} on the test set, demonstrating strong capability in identifying irrelevant results with high accuracy.

\begin{figure}[htbp]
\centering
\includegraphics[width= 0.9\linewidth]{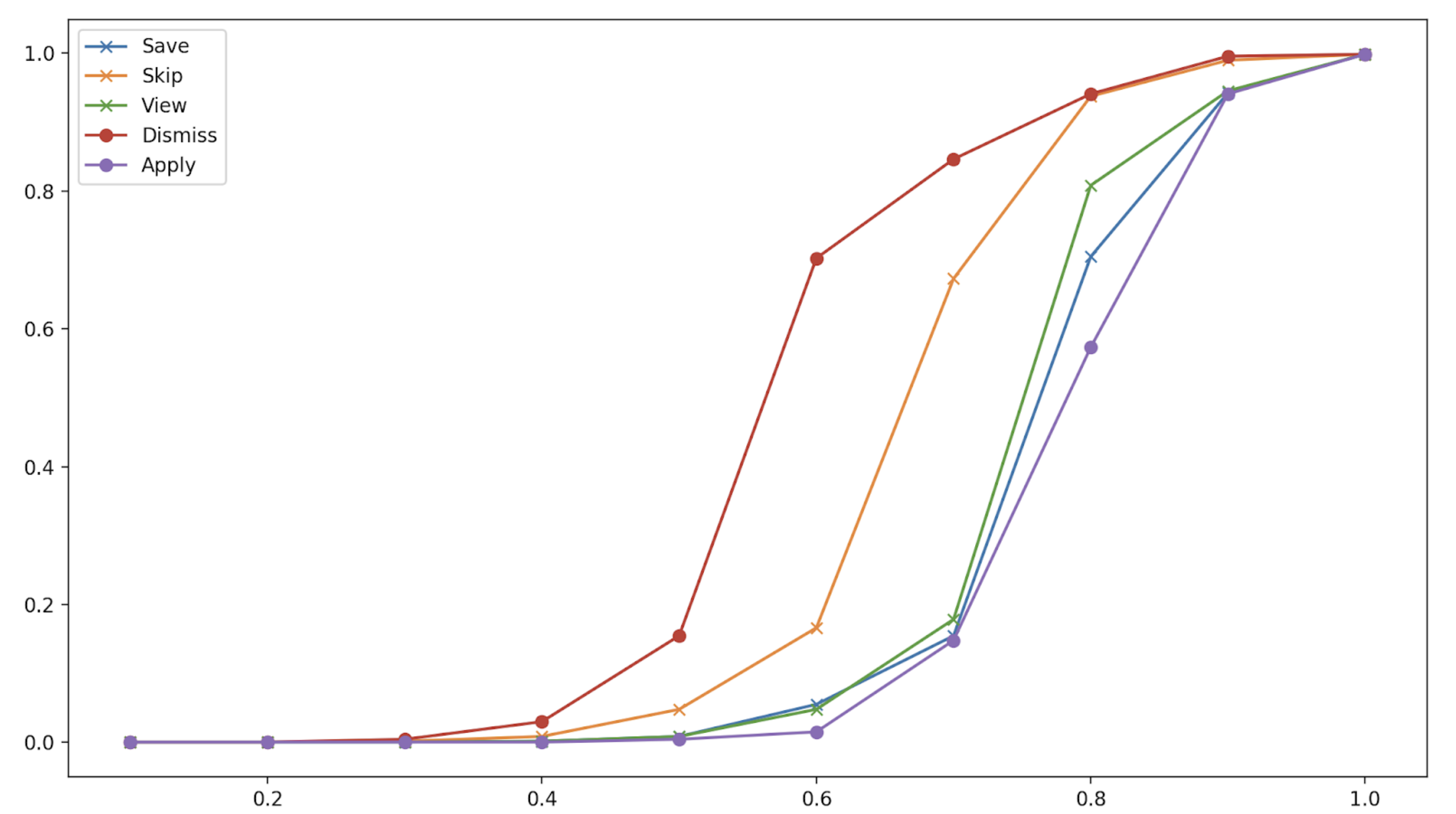}
\caption{Quantiles on the X-axis, p(irrelevantJob) on the Y-axis}
\label{fig:quant}
\end{figure}

To further assess the weak labeler’s effectiveness, we applied it to the search model’s training dataset and analyzed score distributions across various user interactions. As illustrated in Figure~\ref{fig:quant}, we observed that $40\%$ of dismissed jobs received an irrelevance score above $0.7$, while only $20\%$ of applied jobs exceeded a score of $0.6$. This pattern aligns with expectations, reflecting a gradation in relevance across user interactions — from apply, save, view, skip, to dismiss.

We also examined edge cases to validate the model’s predictions. Specifically, we spot-checked jobs that were dismissed but had low irrelevance scores, as well as jobs that were applied to despite having high irrelevance scores. These anomalies aligned with known user behaviors: dismissals can occur for various reasons unrelated to relevance, and some users apply broadly to multiple jobs regardless of fit.

\subsubsection{Search Model Evaluation}
We evaluated the model following the approach in Section~\ref{sec:eval}, using both the original labels (user interactions) and the weak labeler’s outputs. As shown in Table~\ref{tab:performance_metrics}, the weak supervision approach significantly improved NDCG@10 when evaluated against these probabilistic labels, indicating successful knowledge transfer to the ranking model. This improvement came with only a minor drop in NDCG@10 based on user engagement labels, suggesting that the model retained most of its original performance while incorporating the new signals. Additionally, the observed increase in query-job feature importance highlights that the ranking model learned stronger query-document relevance patterns, aligning with the design of the LFs focused on query-job semantic matching.

\begin{table}[htbp]
\centering
\caption{Weakly Supervised Model Performance Metrics}
\label{tab:performance_metrics}
\begin{tabular}{lc}
\toprule
\textbf{Metric} & \textbf{Relative Change} \\\midrule
NDCG@10 (Original Labels) &  $-1\%$ to $-2\%$ \\
NDCG@10 (Weak Labels) & $+34\%$ to $+42\%$ \\
Query-Job Feature Importance & $+15\%$ to $+30\%$ \\\midrule
Rule-Based Mismatch Rates & $-9\%$ to $-15\%$ \\
Job Sessions & $+0.8\%$ \\
Positive Recruiter Ratings & $+11\%$ \\
\bottomrule
\end{tabular}
\end{table}

\subsection{Online Results}
We deployed multiple versions of the weakly supervised ranking model into our search stack, using the same weak labeling model across all variants while varying only the relabeling approach (Section~\ref{sec:relabel}). Performance was measured using proxy indicators for search quality (rule-based heuristics), user engagement (job sessions), and down-funnel outcomes (recruiter interactions).

Variant \textbf{R1} improved relevance and engagement but negatively impacted revenue. \textbf{R2} further improved relevance but reduced job applications and increased dismissals, likely due to $y_p < y_{dismiss}$ lowering the importance of dismissal-related features in the model. \textbf{R3} addressed \textbf{R1}'s revenue issues and achieved our business objective of improved search precision, as shown in Table~\ref{tab:performance_metrics}. Specifically, it reduced rule-based mismatch rates, and increased job sessions through more user engagement with job alerts. Job alert quality is highly dependent on the top $k$ results, suggesting improved relevance at higher-ranking positions. An increase in positive recruiter ratings also indicates that more users applied to jobs that they were a good fit for.
\section{Conclusions and Future Work}
\label{sec:conclusion}
We highlighted the importance of relevance-labeled datasets over solely relying on user activity logs for training ranking models. Given the time and resource demands of manual labeling, weak supervision offers a scalable alternative by incorporating subject matter expertise and external knowledge sources into the labeling process. In this work, we detailed the design of our end-to-end weak supervision system and shared results from both offline experiments and a successful online deployment.

Future improvements include making our LFs serveable, as current online features only partially align with offline LFs. Enhancing the weak labeler itself is another avenue -- either by improving model performance with fewer assumptions or by reducing reliance on a golden dataset. We also plan to explore using LLMs as LFs~\cite{kojima2022large}. While generalist LLMs still struggle with complex search relevance tasks~\cite{liu2023chatgpt}, they could be effective for simpler labeling tasks, potentially replacing parts of the weak supervision pipeline. Emerging research further explores LLMs as judges, annotators, or reasoning agents~\cite{hsieh2023distilling}, offering promising directions for future work.

\bibliography{paper}

\appendix

\end{document}